\documentclass[a4paper, 12pt, reqno]{amsart}

\usepackage{review}
\usepackage[T1]{fontenc}
\usepackage[utf8]{inputenc}
\usepackage{hyperref}
\begin{document}

\title{Hierarchical Structure in the Trace Formula}

\abstract
Guztwiller's Trace Formula is central to the semiclassical theory of quantum energy levels and spectral statistics in classically chaotic systems.  Motivated by recent developments in Random Matrix Theory and Number Theory, we elucidate a hierarchical structure in the way periodic orbits contribute to the Trace Formula that has implications for the value distribution of spectral determinants in quantum chaotic systems.  
\endabstract

\author{J. P. Keating}
\email{keating@maths.ox.ac.uk}
\address{Mathematical Institute, University of Oxford, Oxford, OX2 6GG, United Kingdom}

\maketitle

\begin{center}
{\em Dedicated to Michael Berry, to mark his 80$^{th }$ birthday.} 
\end{center}



\section{Introduction}\label{sec:introduction}

The trace formula, derived by Martin Gutzwiller, connects quantum energy levels and classical periodic orbits \cite{Gutz}. It is central to the semiclassical theory of energy levels and spectral statistics in the field of Quantum Chaos. For example, it has played an important role in many of Michael Berry's most significant papers on the theory of quantum chaotic systems, including his work with Michael Tabor on spectral statistics in classically integrable systems \cite{BT}, and in his work on the semiclassical theory of spectral rigidity in classically chaotic systems \cite{Berry85}.  The seminal ideas he introduced are beautifully reviewed in his Bakerian Lecture \cite{BerryBak}. 

When I first arrived in Bristol in 1985, as Michael's doctoral student, questions about the trace formula were a major focus of attention (alongside questions about the newly discovered Geometric Phase).  Issues relating to the convergence of the sum over periodic orbits, its accuracy, and its applicability in various situations, were hotly debated.  Michael played an important role in resolving them.  We thought together about summing over repetitions of a single primitive orbit \cite{KB87}, and resumming the pseudo-orbits that collectively represent the quantum spectral determinant when $\hbar\to 0$ \cite{BK90, K92, BK92}.  Michael encouraged me to derive a trace formula for quantum cat maps, where it is an identity \cite{K91}. It was a tremendously exciting time and I learned a great deal from him, John Hannay, Jonathan Robbins, and from the many visitors who came to Bristol. 

One idea of Michael's that particularly inspired me was his observation that the trace formula for quantum chaotic systems bears a striking similarity to the {\em explicit formula} relating the zeros of the Riemann zeta-function to the primes \cite{Berry86}.  I still vividly recall reading for the first time his remarkable and prescient paper in a laundrette in Redland while my clothes were being washed.  This analogy, and in particular the Riemann-Siegel formula for the zeta-function, inspired us to develop the Riemann-Siegel Lookalike formula \cite{BK90} and then to derive its resummation offspring \cite{K92, BK92}.  It also motivated me to think about how subtle correlations between the primes, as captured by the  Hardy-Littlewood conjecture, conspire, by virtue of the explicit formula, to give rise to random-matrix statistics in the zeros of the zeta function (specifically, to statistics that coincide with those of the eigenvalues of random unitary matrices) when the appropriate limits are taken \cite{montgomery73, KVar1, KVar2, bogkea95, bogkea96}. This in turn led people to consider what the corresponding correlations would have to be between classical periodic orbits in order that the trace formula reproduces the full predictions of Random Matrix Theory \cite{arg}. Michael's formula for the number variance of the Riemann zeros \cite{BerryVar} opened my eyes to the possibility of cross-pollination between Quantum Chaos and Number Theory, leading to precise formula for the zero statistics \cite{bogkeacorr} and for the moments of the zeta function on its critical line \cite{keasna00a, keasna00b, cfkrs03, cfkrs05, conkea15a, conkea15b,  conkea15c, conkea16, conkea19}.  It was a tremendously exciting time and a wonderfully stimulating way to start one's research career.  I learned from Michael to focus on ideas, irrespective of whether they came from Mathematics or Physics, and his group was second to none as an environment in which to explore the boundary between these fields.       

As a small contribution to the celebration of Michael's $80^{\rm th}$ birthday, I would, in this short note, like to draw attention to a hierarchical structure in the way periodic orbits contribute to the trace formula that was not apparent when we were thinking about these matters thirty years ago. We have only recently begun to understand this structure, and it consequences, in the context of the explicit formula relating the zeros of the Riemann zeta-function to the primes and in the corresponding formulae  for random matrices, and my purpose here is to transfer what we have learned to semiclassical periodic orbit expressions.  The fact that this extension to the semiclassical theory of quantum chaos is inspired by Michael's suggested analogy with the primes is but another example of the lasting impact his ideas have had. 

The results concerning the explicit formula relating the zeros of the Riemann zeta-function to the primes and the corresponding formulae for random matrices that inspire the calculations reported here have recently been reviewed at some length \cite{baikea22}.  I refer readers to that review for appropriate context and details.

\section{Spectral determinants and trace formulae}\label{sec:det}

Consider a system with quantum Hamiltonian $H$ and energy levels $E_n$.  The spectral determinant may be represented formally by
\begin{equation}
\Delta(E)=\det (E-H)=\prod_n (E-E_n).
\end{equation} 
In practice, one may need to regularise the determinant and the product, but this is a standard procedure that does not influence the calculations to be described below and so I will not expand further on this aspect of the theory -- see, for example, \cite{KeaSieb} for details.

We shall be interested in
\begin{equation}
\log\Delta(E)=\log\det (E-H)={\rm Tr}\log (E-H).
\end{equation} 
This is the integral of the energy-dependent Green function and so, using the Gutzwiller trace formula, semiclassically (i.e.~as $\hbar\to 0$),
\begin{equation}\label{eq:trace2}
\log\Delta(E)\sim\sum_pA_p{\rm e}^{{\rm i}S_p(E)/\hbar},
\end{equation} 
where the sum runs over classical periodic orbits, indexed $p$, with stability amplitude $A_p$ and action $S_p$ \cite{Gutz, BK90, K92}.  We shall restrict our discussion to quantum chaotic systems that are not time-reversal symmetric, so to cases where we expect the spectral statistics to be modelled by the Gaussian (GUE) or Circular (CUE) Unitary Ensembles of Random Matrix Theory. 

In the case of the Riemann zeta-function, which is defined for ${\rm Re}s>1$ by \cite{edwards74, titchmarsh86}
\begin{equation}
\zeta(s)\coloneqq\sum_{n=1}^\infty\frac{1}{n^s}= \prod_p\left(1-\frac{1}{p^s}\right)^{-1},
\end{equation}
where $p$ now labels the primes, the corresponding formula is
\begin{equation}\label{eq:zeta}
\log\zeta\left(1/2+{\rm i}E\right)=\sum_p\sum_{r=1}^\infty\frac{1}{p^{r/2}}{\rm e}^{-{\rm i}Er\log p}.
\end{equation}

The sums in \eqref{eq:trace2} and \eqref{eq:zeta} do not convergence. Indeed, the left hand sides are infinite at the positions of the energy levels $E_n$  and at the non-trivial zeros of the zeta function respectively. However, when  the right hand side of  \eqref{eq:trace2} is restricted smoothly to periodic orbits with periods $T_p\lesssim T_H\coloneqq 2\pi\hbar\bar{d}(E)$, where $\bar{d}(E)$ is the mean density of states, or when the sum on the right hand side of  \eqref{eq:zeta} is restricted smoothly to $r\log p\lesssim\log\left(E/2\pi\right)$, the sums do approximate the respective left hand sides in averages over $E$ or in a measure-theoretic sense.  The timescale $T_H$ is often referred to as the {\em Heisenberg time}.  For instance, the moments of $\log\zeta\left(1/2+{\rm i}E\right)$ can be computed by analysing the sum on the right hand side of  \eqref{eq:zeta} restricted smoothly to $r\log p\lesssim\log\left(E/2\pi\right)$ \cite{titchmarsh86, baikea22}. And the restricted sum does approximate $\log\zeta\left(1/2+{\rm i}E\right)$ point-wise for a set of values of $E$ of measure close to one.  Essentially, it is a good approximation except for $E$ close to a zero of the zeta function \cite{baikea22}.   

Restricting the sum on the right hand side of \eqref{eq:trace2}  smoothly to periodic orbits with periods $T_p\lesssim T_H$, modelling the actions $S_p$ as independent random variables -- this is equivalent to what is often termed the {\em diagonal approximation} -- and using the Hannay-Ozorio de Almeida sum rule \cite{HO}, which implies that when $T_2\gg T_1$
\begin{equation}\label{eq:HO}
\sum_{T_1\le T_p\le T_2}
|A_p|^2 \approx 
\int_{T_1}^{T_2}\frac{{\rm d}T}{T}
=\log\left(\frac{T_2}{T_1}\right),
\end{equation}
the moments of $\log\Delta(E)$ can be computed semiclassically, as $\hbar\to 0$ to be those of a normal distribution with variance $\frac{1}{2}\log \left(T_H/T_0\right)$, where $T_0$ is a time-scale characterising the short-time classical dynamics (and so is independent of $\hbar$)\footnote{For example, $T_0$ might be the period of the shortest periodic orbit.}.  Put another way, 
\begin{equation}
\frac{\log\Delta(E)}{\sqrt{\frac{1}{2}\log \left(T_H/T_0\right)}}
\end{equation}
convergences semiclassically to a standard complex normal random variable (i.e.~to a complex random variable with real and imaginary parts that are independent normal random variables with zero mean and unit variance).   This is consistent with what can be proved for the characteristic polynomials of random matrices drawn from the GUE \cite{CL} or CUE \cite{keasna00a} of Random Matrix Theory.

It is worth remarking in passing that this central limit theorem implies that $\Delta(E)$ behaves highly erratically in the semiclassical limit, in that it implies that for any fixed $X$, no matter how big, the probability that $|\Delta(E)|$ takes a value $>X$ tends to $1/2$, and the probability that $|\Delta(E)|$ takes a value $<1/X$ tends to $1/2$. 

The corresponding central limit theorem in the case of the Riemann zeta-function, proved by Selberg, is that 
\begin{equation}
\frac{\log\zeta\left(1/2+{\rm i}E\right)}{\sqrt{\frac{1}{2}\log \log E}}
\end{equation} 
convergences to a standard complex normal random variable when $E\to\infty$ \cite{titchmarsh86}.   Again, this is consistent with what can be proved for the characteristic polynomials of random matrices drawn from the CUE or GUE of Random Matrix Theory. 

Extending this approach to compute the covariance -- that is, applying the trace formula and invoking the diagonal approximation -- we have that for $\epsilon\ll E$
\begin{equation}
\left< \log|D(E)|\log|D(E+\epsilon)|\right>_E\sim \frac{1}{2}{\rm Re}\sum_{T_p\lesssim T_H}|A_p|^2{\rm e}^{{\rm i}\epsilon T_p/\hbar},
\end{equation} 
where $< \dots>_E$ denotes a local average around $E$ over a range that is classically small but semiclassically large. Hence, applying the sum rule \eqref{eq:HO}
\begin{equation}
\left< \log|D(E)|\log|D(E+\epsilon)|\right>_E\sim \begin{cases}
			\frac{1}{2}\log \left(T_H/T_0\right), & \text{if $\epsilon\bar{d}\ll 1$}\\
            -\frac{1}{2}\log\left(\epsilon T_0/\hbar \right), & \text{if $\epsilon\bar{d}\gg 1$ and $\epsilon T_0/\hbar \ll 1$}
		 \end{cases}   .
\end{equation}
It follows that $\log\Delta(E)$ behaves like a logarithmically correlated Gaussian random variable, and so is similar to a wide class of mathematical objects that have recently been of considerable interest \cite{fyobou08}.  This is also the case for the Riemann zeta-function and for characteristic polynomials of random matrices.  See \cite{baikea22} for a review. 

My purpose here is examine one particular aspect of the theory of  logarithmically correlated Gaussian random fields in the context of the semiclassical theory of quantum chaotic systems.  This manifests itself as a hierarchical structure in the trace formula \eqref{eq:trace2} that has, as far as I am aware, not previously been discussed.

\section{Hierarchical organisation in the trace formulae}\label{sec:tree}

We shall examine the trace formula \eqref{eq:trace2} close to the energy $E$. In order to focus on a small energy window around this energy, let us write
\begin{equation}\label{eq:trace}
\log|\Delta(E+\epsilon)|\sim{\rm Re}\sum_pA_p{\rm e}^{{\rm i}S_p(E)/\hbar+{\rm i}\epsilon T_p/\hbar},
\end{equation} 
with $0\le\epsilon <\epsilon_0$.

The number of energy levels in this window is $\sim\epsilon_0\bar{d}(E)$ and to approximate $\log|\Delta(E+\epsilon)|$ we expect to sum over periodic orbits with $T_p\lesssim T_H$.  One can think of $\log|\Delta(E+\epsilon)|$ as being roughly constant between consecutive energy levels, at which it has (logarithmic) singularities: obviously $|\Delta(E+\epsilon)|$ itself takes a range of values between consecutive zeroes, but its logarithm is relatively flat.   One can therefore ask how the periodic periodic orbits sum up to give the  $\sim\epsilon_0\bar{d}(E)$ values taken by $\log|\Delta(E+\epsilon)|$ within the energy window.  

To explore this question, we split the sum up into dyadic intervals, labelled by $k$, containing periodic orbits with periods $T_0 2^k\le T_p<T_02^{k+1}$:
\begin{equation}\label{eq:trace}
\log|\Delta(E+\epsilon)|\sim{\rm Re}\sum_{k=0}^{k_{\rm max}}\sum_{T_0 2^k\le T_p<T_02^{k+1}}A_p{\rm e}^{{\rm i}S_p(E)/\hbar+{\rm i}\epsilon T_p/\hbar}.
\end{equation}
Here $T_02^{k_{\rm max}+1}\sim T_H$, i.e.
\begin{equation}
k_{\rm max}\sim\log_2\frac{T_H}{2T_0}
\end{equation} 
It will be clear as we proceed that dyadic intervals have been chosen solely for illustrative purposes; one could equally well split the sum up according to $T_0 X^k\le T_p<T_0X^{k+1}$ for any convenient $X$.

Setting
\begin{equation}
Y_k(E; \epsilon)\coloneqq {\rm Re}\sum_{T_0 2^k\le T_p<T_02^{k+1}}A_p{\rm e}^{{\rm i}S_p(E)/\hbar+{\rm i}\epsilon T_p/\hbar},
\end{equation}
it is clear that the energy averages of $Y_k(E; \epsilon)$ and $(Y_k(E; \epsilon))^2$ are given semiclassically by
\begin{equation}
<Y_k(E; \epsilon)>_E\approx 0 
\end{equation}
and
\begin{align}
<(Y_k(E; \epsilon))^2>_E&\approx\frac{1}{2}\sum_{T_0 2^k\le T_p<T_02^{k+1}}|A_p|^2\nonumber\\
       &\approx\frac{1}{2}\int_{T_0 2^k}^{T_02^{k+1}}\frac{{\rm d}T}{T}\nonumber\\
       &=\frac{1}{2}\log 2,
\end{align}
using the diagonal approximation and the Hannay-Ozorio de Almeida sum rule \eqref{eq:HO}.   The higher moments can be calculated in the same way to be those of a Gaussian with zero mean and variance $\frac{1}{2}\log 2$, so as $\hbar\to 0$
\begin{equation}
Y_k(E; \epsilon))\rightarrow\mathcal{N}(0,\frac{1}{2}\log 2).
\end{equation}  

In the same way the diagonal approximation can be used to establish that when $k\ne l$
\begin{align}
<Y_k(E; \epsilon)Y_{l}(E; \epsilon^\prime)>_E&\sim\frac{1}{2}
{\rm Re}\sum_{T_0 2^k\le T_p<T_02^{k+1}}
\sum_{T_0 2^l\le T_q<T_02^{l+1}}A_p(A_q)^*
{\rm e}^{{\rm i}(S_p(E)-S_q(E))/\hbar+{\rm i}(\epsilon T_p-\epsilon^\prime T_q)/\hbar}\nonumber\\
       &\approx 0
\end{align}
because there are no diagonal terms in this case, and that
\begin{align}\label{eq:cov}
<Y_k(E; \epsilon)Y_{k}(E; \epsilon^\prime)>_E&\sim\frac{1}{2}
{\rm Re}\sum_{T_0 2^k\le T_p<T_02^{k+1}}
\sum_{T_0 2^k\le T_q<T_02^{k+1}}A_p(A_q)^*
{\rm e}^{{\rm i}(S_p(E)-S_q(E))/\hbar+{\rm i}(\epsilon T_p-\epsilon^\prime T_q)/\hbar}\nonumber\\
       &\approx \frac{1}{2}{\rm Re}\sum_{T_0 2^k\le T_p<T_02^{k+1}}|A_p|^2{\rm e}^{{\rm i}(\epsilon -\epsilon^\prime )T_p/\hbar}\nonumber\\
       &\approx \frac{1}{2}{\rm Re}\int_{T_0 2^k}^{T_02^{k+1}}\frac{{\rm d}T}{T}{\rm e}^{{\rm i}(\epsilon -\epsilon^\prime )T/\hbar}\nonumber\\
       &\approx \begin{cases}
			\frac{1}{2}\log 2, & \text{if $|\epsilon-\epsilon^\prime|2^kT_0/\hbar\ll 1$}\\
            0, & \text{if $|\epsilon-\epsilon^\prime|2^kT_0/\hbar\gg 1$}
		 \end{cases}.
\end{align}

The condition for the covariance to be non-zero is equivalent to
\begin{equation}\label{cond1}
k\ll \log_2\frac{\hbar}{T_0|\epsilon-\epsilon^\prime|}
\end{equation}
or
\begin{equation}\label{cond2}
|\epsilon-\epsilon^\prime|\ll\frac{\hbar}{T_0}2^{-k}
\end{equation}
Setting 
\begin{equation}
\epsilon=\frac{x}{\bar{d}(E)},
\end{equation}
\eqref{cond1} becomes
\begin{align}
k&\ll \log_2\frac{T_H}{T_0|x-x^\prime|}\nonumber\\
&=\log_2\frac{2^{k_{\rm max}+1}}{|x-x^\prime|}\nonumber\\
&=k_{\rm max}+1-\log_2|x-x^\prime|
\end{align}
and \eqref{cond2} becomes
\begin{align}
|x-x^\prime|&\ll \frac{T_H}{T_0}2^{-k}\nonumber\\
&=2^{k_{\rm max}-k+1}
\end{align}

The overall picture that emerges is that the periodic orbit contributions to the trace formula \eqref{eq:trace} can be grouped together so that
\begin{equation}\label{eq:traceY}
\log|\Delta(E+\epsilon)|\sim{\rm Re}\sum_{k=0}^{k_{\rm max}}Y_k(E; \epsilon),
\end{equation}
where the $\sim\log_2\frac{T_H}{2T_0}$ summands $Y_k(E; \epsilon)$ behave semiclassically like normal random variables with mean zero and variance $\frac{1}{2}\log 2$.  This is, of course, consistent with the fact that $\log\Delta(E)$ has a normal distribution with zero mean and variance $\frac{1}{2}\log \left(T_H/T_0\right)$.  Crucially, however, the summands at different points $\epsilon$ and $\epsilon^\prime$ are not independent of each other. Rather, they are, roughly speaking, perfectly correlated for $k\lesssim k_{\rm LCA}(|\epsilon-\epsilon^\prime|)$, where
\begin{equation}
k_{\rm LCA}(|\epsilon-\epsilon^\prime|)=\log_2\frac{\hbar}{T_0|\epsilon-\epsilon^\prime|},
\end{equation}
or, when energy differences are measured on the scale of the mean level separation ($\epsilon=x/\bar{d}(E)$)
\begin{equation}
k_{\rm LCA}(|\epsilon-\epsilon^\prime|)=k_{\rm max}+1-\log_2|x-x^\prime|.
\end{equation}
That is, the contributions to the trace formula at $\epsilon$ and $\epsilon^\prime$ coincide when $k\lesssim k_{\rm LCA}(|\epsilon-\epsilon^\prime|)$, but then become uncorrelated.  The closer  $\epsilon$ is to $\epsilon^\prime$, the larger $k_{\rm LCA}(|\epsilon-\epsilon^\prime|)$ is, and so the more terms in the periodic orbit sum are correlated, when these are aggregated into dyadic intervals.

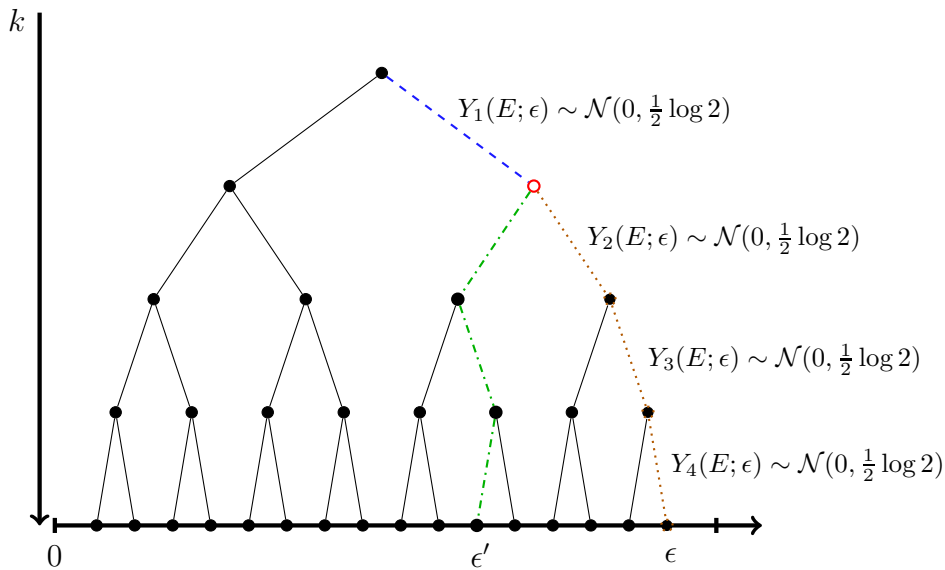
\begin{figure}[htb]
  \centering
  \begin{tikzpicture}[
      level distance=1.5cm,
      level 1/.style={sibling distance=4cm},
      level 2/.style={sibling distance=2cm},
      level 3/.style={sibling distance=1cm},
      level 4/.style={sibling distance=.5cm},
      every node/.style = {shape=circle, inner sep=1.5pt, draw, align=center, fill=black}]
    \draw[->,ultra thick] (-4.3,-6)--(5,-6);
    \draw[<-,ultra thick] (-4.5,-6)--(-4.5,0.8);
    \draw[-,ultra thick] (4.4,-5.9)--(4.4,-6.1);
    \draw[-,ultra thick] (-4.3,-5.9)--(-4.3,-6.1);
    \node[draw=none,fill=none] at (-4.3,-6.4) {$0$};
    \node[draw=none,fill=none] at (-4.8,.7) {$k$};
    \node[draw=none,fill=none] at (2.8,-.5) {\footnotesize{$Y_1(E;\epsilon)\sim\cN(0,\frac{1}{2}\log 2)$}};
    \node[draw=none,fill=none] at (4.5,-2.2) {\footnotesize{$Y_2(E;\epsilon)\sim\cN(0,\frac{1}{2}\log 2)$}};
    \node[draw=none,fill=none] at (5.3,-3.8) {\footnotesize{$Y_3(E;\epsilon)\sim\cN(0,\frac{1}{2}\log 2)$}};
    \node[draw=none,fill=none] at (5.6,-5.2) {\footnotesize{$Y_4(E;\epsilon)\sim\cN(0,\frac{1}{2}\log 2)$}};
    \node[draw=none,fill=none] at (3.8,-6.4) {$\epsilon$};
    \node[draw=none,fill=none] at (1.3,-6.4) {$\epsilon^\prime$};
    \node {}
    child {node {}
      child {node {}
        child {node {}
          child {node {}}
          child {node {}}
        }
        child {node {}
          child {node {}}
          child {node {}}
        }
      }
      child {node {}
        child {node {}
          child {node {}}
          child {node {}}
        }
        child {node {}
          child {node {}}
          child {node {}}
        }
      }
    }
    child[blue!90!white,dashed,thick] {node[solid,draw=red,circle,fill=white,thick] {}
      child[green!70!black,dashdotted,thick] {node[solid,draw=black,circle] {}
        child[black,solid,thin] {node[solid,draw=black,circle] {}
          child {node[solid,draw=black,circle] {}}
          child {node[solid,draw=black,circle] {}}
        }
        child {node[solid,draw=black,circle] {}
          child {node[solid,draw=black,circle] {}}
          child[black,solid,thin] {node[solid,draw=black,circle] {}}
        }
      }
      child[orange!70!black,dotted,thick] {node {}
        child[black,solid,thin] {node {}
          child {node {}}
          child {node {}}
        }
        child {node {}
          child[black,solid,thin] {node {}}
          child {node {}}
        }
      }
    };
  \end{tikzpicture}
  \caption[Tree structure.]{An example of paths on a binary tree of depth $n=4$, from root to leaves $\epsilon$ and $\epsilon^\prime$.  Some weightings $Y_j(l)$ are highlighted, where $Y_j(l)\sim \cN(0,\frac{1}{2}\log 2)$. The last common ancestor of leaves $\epsilon, \epsilon^\prime$ is illustrated by the `hollow' (red) node and occurs at level $1$.}\label{fig:binary_tree}
\end{figure}

I illustrate this behaviour in figure~\ref{fig:binary_tree}.  The contributions to \eqref{eq:traceY} can be arranged to form a binary tree. The leaves at level $k=k_{\rm max}$ correspond to energies between adjacent energy levels.  At any energy $\epsilon$ the terms in the sum can be  thought of as sitting on the edges of the unique path through the tree starting at the root and ending at the corresponding leaf. Different leaves have different associated paths. They may have some initial  edges in common, but the paths eventually diverge at a {\em Last Common Ancestor} at level $ k_{\rm LCA}(|\epsilon-\epsilon^\prime|)$.

The hierarchical structure represented in figure~\ref{fig:binary_tree} represents the principal observation I seek to explain here. I believe it casts new light on the way periodic orbits contribute, via the trace formula, to characterising the semiclassical statistical properties of spectral functions in quantum chaotic systems. 

\section{Connections}\label{sec:connections}

It is worth remarking that the hierarchical structure described in Section \ref{sec:tree} can be proved in the case of the explicit formula \eqref{eq:zeta} for the Riemann zeta-function; see \cite{baikea22}  for an overview. Moreover, it can be proved for the corresponding formula in Random Matrix Theory. Let $A$ be a random $n\times n$ unitary matrix, drawn from the Circulate Unitary Ensemble, and let
\begin{equation}
P(A,\theta)=\det(I-A{\rm e}^{-{\rm i}\theta})
\end{equation} 
denote its characteristic polynomial.  Then
\begin{equation}\label{eq:fourier}
\log|P(A,\theta)|=-{\rm Re}\sum_{m=1}^\infty\frac{{\rm Tr}A^m}{m}{\rm e}^{-{\rm i}m\theta},
\end{equation}
where the convergence properties of the sum are subtle \cite{hugkeaoco01}, but may be understood in a distributional sense.  It can be shown that when $n\to\infty$, the value distribution of $\log|P(A,\theta)|/\sqrt{\frac{1}{2}\log n}$ converges to a standard normal and that at different values of $\theta$, $\log|P(A,\theta)|$ is log-correlated \cite{baikea22}: 
\begin{equation}\label{eq:vn_corr}
  \mathbb{E}[\log|P(A,\theta_1)|\log|P(A,\theta_2)|]\approx
  \begin{dcases}
    -\frac{1}{2}\log|\theta_1-\theta_2|, & \text{for }\frac{1}{n}\ll |\theta_1-\theta_2|\ll 1\\
    \frac{1}{2}\log n, & \text{for }|\theta_1-\theta_2|\ll \frac{1}{n}.
  \end{dcases}
\end{equation}
Splitting the sum \eqref{eq:fourier} up into diadic intervals, it can be proved that the appropriately aggregated summands have precisely the hierarchical structure described in Section \ref{sec:tree}. I again refer to the recent review \cite{baikea22} for further details.

If one models the set of variables $Y_k(E; \epsilon)$ associated to the trace formula, and their analogues for the zeta function and the sum \eqref{eq:fourier}, as having {\em precisely} a normal distribution and being {\em precisely} log-correlated, then these problems all map onto the much-studied {\em branching random walk}.   The branching random walk can be analysed in considerable detail \cite{baikea20b}, and the results may then be expected to hold for the spectral determinants of Schr\"odinger operators in quantum chaotic systems, the zeta function, and the characteristic polynomials of random matrices. In the latter two cases, this has been the motivation for a good deal of recent research \cite{fyohiakea12, fyokea14, web15, arg16, argbelbou17, paqzei17, fyognukea18, naj18, baikea19, abbrs19, argbourad20, assbaikea20, asskea20, forkea20, keawon20, niksakweb20}. It would appear that the case of spectral determinants would also merit further investigation in this context. For example, it is natural to conjecture that, like the zeta function and the characteristic polynomials of random matrices, they converge in the semiclassical limit to the Gaussian Multiplicative Chaos Measure \cite{rhovar14}, and that their extreme value statistics should fall into the class of log-correlated Gaussian fields, an example of which is the Branching Random Walk.

\section{Acknowledgements}\label{sec:acknowledgements}

This work was supported by ERC Advanced Grant 740900 (LogCorRM).   I am grateful to Louis-Pierre Arguin, Emma Bailey, Yan Fyodorov and Ofer Zeitouni for helpful discussions.   


\end{document}